\documentclass[11pt,a4paper]{article}
\usepackage[utf8]{inputenc}
\usepackage[english]{babel}
\usepackage{amsmath}
\usepackage{amsfonts}
\usepackage{amssymb}
\usepackage{graphicx}
\usepackage{natbib}
\usepackage{hyperref}
\usepackage{fancyhdr}
\usepackage[left=2cm,right=2cm,top=2cm,bottom=2cm]{geometry}
\author{Kasper Kansanen$^{1,\star}$, Petteri Packalen$^2$, Matti Maltamo$^2$, and Lauri Mehtätalo$^1$}
\title{Horvitz--Thompson-like estimation with distance-based detection probabilities for circular plot sampling of forests}

\begin{document}
$\quad$\\
$\quad$\\
$\quad$\\
$\quad$\\
$\quad$\\
$\quad$\\
$\quad$\\
$\quad$\\
$\quad$\\
$\quad$\\
$\quad$\\
$\quad$\\
$\quad$\\
$\quad$\\
$\quad$\\
\noindent \textbf{DISCLAIMER:} This is the pre-peer reviewed version of the article, which has been published in Biometrics at \href{https://doi.org/10.1111/biom.13312}{doi.org/10.1111/biom.13312}. This preprint may be used for non-commercial purposes in accordance with Wiley Terms and Conditions for Use of Self-Archived Versions.
\newpage

\maketitle
\thispagestyle{fancy}
\fancyhf{}
\fancyfoot[L]{\begin{small}\emph{
$^\star$ kasper.kansanen@uef.fi\\
$^1$ School of Computing, University of Eastern Finland, Joensuu, Finland\\
$^2$ School of Forest Sciences, University of Eastern Finland, Joensuu, Finland}\end{small}}
\renewcommand{\headrulewidth}{0pt}
\renewcommand{\footrulewidth}{0.4pt}
\renewcommand{\footskip}{15pt}

\begin{abstract}
In circular plot sampling, trees within a given distance from the sample plot location constitute a sample, which is used to infer characteristics of interest for the forest area. If the sample is collected using a technical device located at the sampling point, e.g. a terrestrial laser scanner, all trees of the sample plot cannot be observed because they hide behind each other. We propose a Horvitz--Thompson-like estimator with distance-based detection probabilities derived from stochastic geometry for estimation of population totals such as  stem density and basal area in such situation. We show that our estimator is unbiased for Poisson forests and give estimates of variance and approximate confidence intervals for the estimator, unlike any previous methods. We compare the estimator to two previously published benchmark methods. The comparison is done through a simulation study where several plots are simulated either from field measured data or different marked point processes. The simulations show that the estimator produces lower or comparable error values than the other methods. In the sample plots based on the field measured data the bias is relatively small -- relative mean of errors for stem density, for example, varying from $0.3$ to $2.2$ per cent, depending on the detection condition -- and the empirical coverage probabilities of the approximate confidence intervals are either similar to the nominal levels or conservative. 
\end{abstract}

\section{Introduction}

Circular plot sampling is a commonly used method in forest inventory \citep{GregoireandValentine2007}. In this form of sampling, a location is selected as a centre point of a plot and the surrounding area within a given distance is observed from that point. The objective is to gather information on forest characteristics of interest, such as stem density $N$ (the number of tree stems per unit area) and basal area $G$ (the sum of the the cross-sections of stems at breast height per unit area). We focus on $N$ and $G$ in this study. If the sampling is performed visually by a person, they can move slightly from the sampling point to observe trees that are not visible from the sampling point because of some sort of an obstacle. However, if the sampling is performed with a technical device, such as a laser scanner,  the device cannot usually be moved from its original position. Hence, these sampling methods are susceptible to errors caused by incomplete detection. Lately, terrestrial laser scanning (TLS) has gained popularity as a tool for circular plot sampling and it is topical to consider the detection problems in these sampling situations. 

TLS is a method of close-range remote sensing where a LiDAR scanner is operated on ground level to produce a three-dimensional point cloud of the surroundings. Typically, the scanner remains stationary on a tripod and rotates $360$ degrees in the horizontal plane while the laser scans in the vertical direction in some device dependent opening angle. In forestry, the strength of TLS is that accurate measurements of the forest structure below the canopy can be made. Methods for detecting individual tree objects from the point cloud have been developed \citep[see e.g.][]{Strahleretal2008,Lovelletal2011,Liangetal2012,Raumonenetal2015,Ravagliaetal2017}, meaning that tree level information for forest inventory purposes can be gathered using TLS. In this study, we focus strictly on tree stems (or more specifically, their cross-sections at a certain reference height) and ignore other parts of trees.

TLS data collection can be performed either as a single- or a multi-scan setup. The difference between these two setups is that in single-scan one forest area is scanned only once, whereas in multi-scan the same area is scanned from several different locations with overlapping scanning areas to produce a more accurate point cloud, having echoes from different sides of trees and possibly from all of the trees in the area. However, scanning from several locations is more time consuming and there is the added difficulty of combining the data from different scans. From estimation point of view, several scans in nearby locations is suboptimal compared to single scans from distant locations.

We focus on the single-scan case where individual tree stems have already been detected from the point cloud, meaning that at least locations and diameters at breast height (DBH) of these detected trees are available. Naturally, the locations and DBH will have estimation errors, but we shall not consider this error source. The problem with single-scan TLS that we are focusing on is that not all trees can be detected, which can lead to underestimation of forest characteristics of interest, such as stem density or basal area. There are different possible reasons for not detecting all of the trees.  For example, the tree stems closer to the scanner produce nonvisible areas behind them so that trees further away from the scanner located in the nonvisible areas can not be seen (Figure~\ref{fig:exampleplot}). Undergrowth, low branches or other objects can block the submitted laser pulse. Small trees far away from the scanner can produce very few laser returns. We consider here only the effect of tree stems producing nonvisible areas behind them; generalization to other obstacles is trivial.

Correcting the problem of nondetection in single-scan TLS has garnered some attention. The use of gap probabilities of a Poisson forest has been proposed as a means for correction \citep[e.g.][]{Lovelletal2011}. \citet{Duncansonetal2014} and \citet{Astrupetal2014} proposed models for detection probability based on traditional distance sampling. \citet{Seidelandammer2014} proposed a correction factor based on the nonvisible area of the TLS plot. \citet{Olofssonandolsson2018} used only the area visible to the scanner as a sampling window. \citet{Kuronenetal2019} improved on the work of  \citet{Olofssonandolsson2018} by modifying the visible area based on what we call a \textit{detection condition}, producing a weight for every tree that depends on its DBH and the detection condition: is the tree detected only if it is fully outside of the nonvisible area, or if the centre point is visible, or if any small visible part of stem is enough for detection, or something in between, a partial visibility? The premise for the work of \citet{Kuronenetal2019} was that the estimator of \citet{Olofssonandolsson2018} produced large under- and over-estimation in Poisson forests when the detection condition was either full visibility or any visibility, respectively, and deduced that this was because the area from which trees could be detected in these cases was not the same as the area visible from the scanner. However, \citet{Kuronenetal2019} also found that their estimator has a positive bias in the Poisson process case.

\citet{Kansanenetal2016} proposed estimators for stem density that correct a nondetection problem in the case of individual tree detection from airborne laser scanning (ALS) data. In ALS, unlike TLS, the scanner is operated from an aircraft and the laser measurements are attained from above, mainly from the forest canopy. They assumed that a tree could be left undetected if it was covered by the larger tree crowns. The most promising of the estimators was a Horvitz--Thompson-like estimator where the detection probability for each tree was calculated based on the planar set formed as a union of the projections of the larger tree crowns on ground that was morphologically transformed based on the detection condition. This method requires the maximum radii of detected tree crowns and the locations of their centres.

Here we take a similar approach for calculating detection probabilities in the TLS case, and more generally, any circular plot sampling case with similar problems with incomplete observation. Instead of the size of the tree crowns, we assume that there is a certain hierarchy on trees being left undetected based on their distances from the scanner. The detection condition is taken into account by a tuning parameter that controls a morphological transformation of the area visible, or nonvisible, from the scanning location. The detection probabilities are used in a Horvitz--Thompson-like estimator to produce estimates of a population total of interest, such as stem density or basal area. The main difference between our estimator and the estimator of  \citet{Kuronenetal2019} is that our detection probabilities are distance-based, whereas their construction is area-based. Hence it could be easier to combine our method with other, possibly distance sampling based, methods that take into account some other effects on tree detection from TLS data that depend on the distance from the scanner. The performance  of the estimator is compared with those of \citet{Olofssonandolsson2018} and \citet{Kuronenetal2019} in circular sampling plots that are either fully simulated from point process models or based on field measurements of diameters and locations of all trees of rectangular fixed-area plots. Three detection conditions are considered: full visibility, centre point visibility, and visibility of any part of the stem.

\section{Methodology}

We model the forest as a marked point process \citep[see e.g.][]{Illianetal2008} $M = \{(x_i, d_i, m_i) \}$, where $x_i$ are the locations of the stem centres, generated by some spatial point process, $d_i$ are the DBH of the trees, governed by some distribution, and $m_i$ is a mark of interest related to tree $i$. If we are interested in a stem density estimate $\hat{N}$, then $m_i=1$. If we are interested in total basal area $\hat{G}$, then $m_i$ is the basal area of tree $i$. Other marks, such as volume or biomass, are also possible. We observe $n$ trees from a total of $N$ in some window of interest $W$ (i.e., fixed-area sample plot), which, without loss of generality, is centred at the origin of the plane.

We model the tree stems as discs $B(x_i, d_i/2)$ centred at $x_i$ with radius $d_i/2$. We assume that the trees can be ordered based on their distances from the origin $r_i - d_i/2$, where $r_i$ is the distance of the stem centre from the origin. In other words, the trees are ordered based on the shortest distance to their outer bark. Henceforth we assume that the index $i$ is ordered, $i=1$ being the tree closest to the origin, $i=2$ being the second closest, and so forth. We also assume that no tree disc $B(x_i, d_i/2)$ covers the origin.

\begin{figure}
\centerline{\includegraphics[width=14cm]{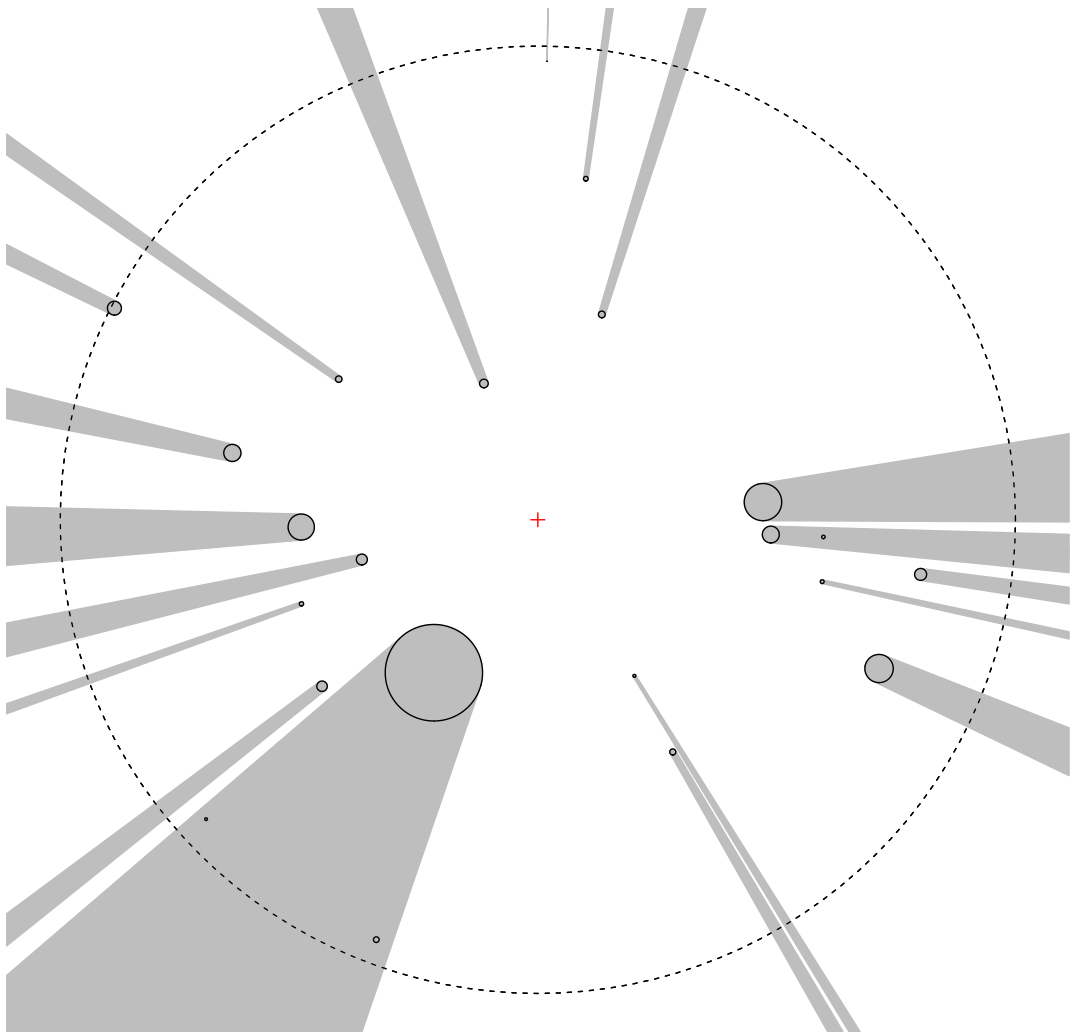}}
\caption{\label{fig:exampleplot} An example of a simulated circular sample plot with the nonvisible areas produced by trees coloured with grey.}
\end{figure}

We assume that the ability to detect tree $i$ is related to the trees closer to the origin than tree $i$. A tree forms a subset of plane $A_i$ that can not be seen from the origin as a union of the stem disc $B(x_i, d_i/2)$ and the area between the two tangent lines of the disc running through the origin (see Figure~\ref{fig:exampleplot}). 

We suggest the following for the detection probability $p$ of tree $i$. If $i=1$ the tree is detected for sure, $p(r_1)=1$. Otherwise,

\begin{equation}\label{eqn: datadete}
p_{\alpha}(r_i, d_i)   =  \left\lbrace\begin{array}{ll}
1-\frac{L[\partial B(o,r_i), (\cup_{j=1}^{i-1}A_j)\ominus B(o, |\alpha| d_i/2)]}{2\pi r_i}, & \alpha<0 \\ 
1-\frac{L[\partial B(o,r_i), \cup_{j=1}^{i-1}A_j]}{2\pi r_i}, & \alpha=0 \\ 
1-\frac{L[\partial B(o,r_i), (\cup_{j=1}^{i-1}A_j)\oplus B(o, \alpha d_i/2)]}{2\pi r_i}, & \alpha>0
\end{array}\right.,
\end{equation}

where $L(\partial  B(o,r), X)$ is the total length of the arcs of a circle with radius $r$ that is inside the set $X$ and $\alpha$ is a tuning parameter that controls the morphological transformations $\ominus$ (erosion) and $\oplus$ (dilation).

\begin{figure}
\centerline{\includegraphics[width=7cm]{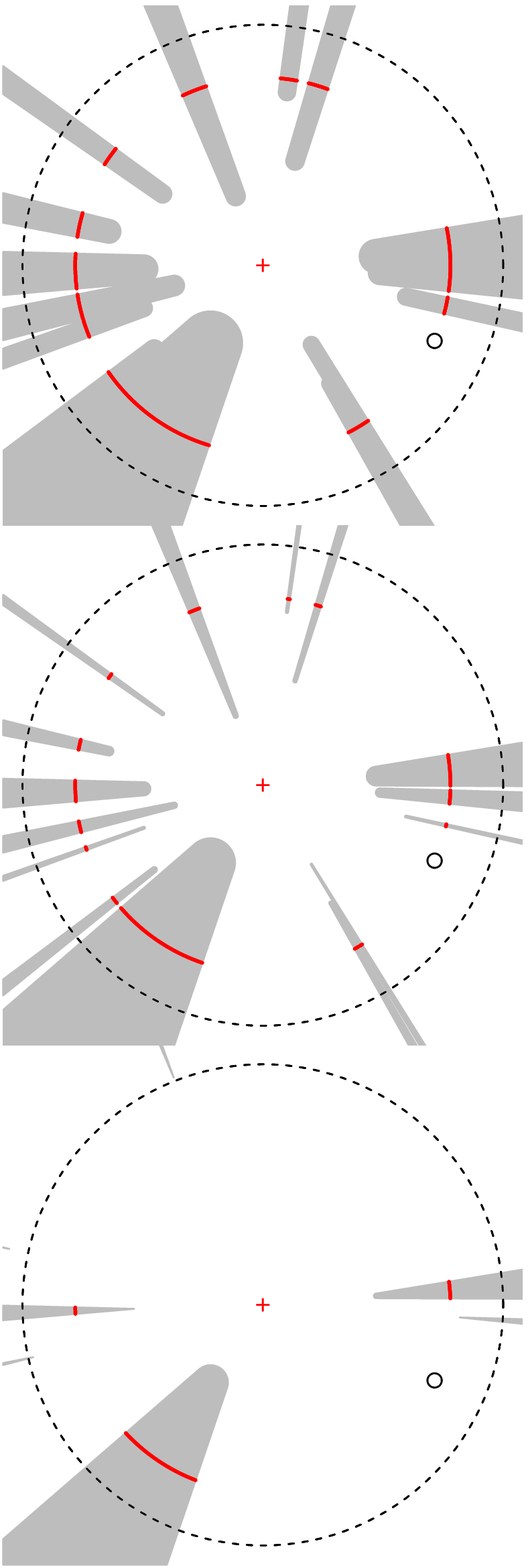}}
\caption{\label{fig:visibilityexample} Examples of areas of nondetection (coloured with grey) related to a tree (the disc) in an example plot. The detection conditions from top to bottom are full visibility, centre point visibility and any visibility. The red circle arcs, coming from a circle with radius corresponding to the distance of the tree centre point from the centre of the plot, are the locations where the tree centre point could be located and left undetected.}
\end{figure}

The reasoning behind the detection probabilities is as follows. If we assume that a point is uniformly distributed on a circle of radius $r$, then the probability that it is located in some certain arc of length $l$ is $l/(2\pi r)$, the proportional length of the arc. Hence, the probability for a tree to be hidden can be calculated based on the lengths of arcs generated by the invisible areas $A_j$, $j<i$. The probability of being detected can then be calculated as a probability of the complement event. The parameter $\alpha$ controls the detection condition. If $\alpha=0$, the centre point of the tree must be visible for detection (centre point visibility detection condition, abbreviated as \textit{centre} in the tables and figures). If $\alpha=-1$, the tree is hidden only if its disc is fully inside the nonvisible area. In other words, any visible part produces detection (any visibility detection condition, abbreviated as \textit{any}). If $\alpha=1$, the tree must be fully visible, or fully outside the union of $A_j$, for detection (full visibility detection condition, abbreviated as \textit{full}). Examples of the nondetectable areas related to these detection conditions are presented in Figure~\ref{fig:visibilityexample}.

With these detection probabilities a Horvitz--Thompson-like estimator for a population total of interest $\tau = \sum_{i=1}^N m_i$ can be formed:

\begin{equation}\label{eqn: HTest}
\widehat{\tau} = \sum_{i=1}^N \frac{m_i I_i}{p_{\alpha}(r_i, d_i)}.
\end{equation}

$I_i$ is an indicator variable that equals $1$ if tree $i$ is detected and $0$ if it is not detected. Hence, in practice only the detection probabilities and marks of detected trees are needed. However, we take into account the invisible areas produced by undetected trees when we calculate the detection probabilities. That is, although a proper detection has not occurred, for example DBH estimation is not possible, these trees increase the nonvisible area and produce a nonzero term $A_j$ to $\cup A_j$ in Equation~\ref{eqn: datadete}. Note that this is not a concern in the case of any visibility detection condition, as the nonvisible areas created by trees that are not detected are fully contained in the nonvisible areas created by the detected trees. 

\subsection{Unbiasedness of the estimator for homogeneous Poisson process}

As a shorthand let us write $p_i$ for the probability of detection for tree $i$ and $S_i$ for the parts of the origin-centred circle with radius $r_i$ that belong to the set from which tree $i$ could not be detected, generated by the trees closer to the origin. Let us write $S_i^c$ for the parts of the circle that do not belong to the aforementioned set, hence, the locations from which tree $i$ could be detected. Let us also write $|S_i|$ for the proportional length of $S_i$, so that $|S_i|+|S_i^c|=1$. Now

\[
E[\widehat{\tau}] = \sum_{i=1}^N \frac{m_i E[I_i]}{p_i}.
\]

Here $I_i$ is an indicator function of tree $i$ belonging to  $S_i^c$. When the homogeneous Poisson process is considered in polar coordinates the angles of the points are uniformly distributed, hence the distribution of a point at a distance $r_i$  is uniform on  a circle of radius $r_i$. From this uniformity it follows that the probability of a point hitting a certain subset of the circle is proportional to the length of that subset. Hence, $E[I_i] = 1\times|S_i^c| + 0\times |S_i| = |S_i^c|$. But now our construction for the probabilities of detection is also $p_i=|S_i^c|$, and it follows that

\[
E[\widehat{\tau}] = \sum_{i=1}^N m_i = \tau,
\]

hence the estimator is unbiased.

\subsection{Estimated variance of the estimator}

When the assumptions related to our Horvitz--Thompson-like estimator are met -- the sequential nature of detection holds, the detection condition is true, and the data is for example generated by the homogeneous Poisson process -- the detection probabilities we approximate in Equation~\ref{eqn: datadete} are the true detection probabilities and hence our estimator coincides with the Horvitz--Thompson estimator.  An unbiased estimator for the variance of the Horvitz--Thompson estimator is known  \citep[see e.g. ][p. 70]{Thompson2012}:

\[
\widehat{var}(\widehat{\tau}) = \sum_{i=1}^n \left(\frac{1}{p_i^2} -  \frac{1}{p_i}\right)m_i^2 + 2\sum_{i=1}^n\sum_{j>i}\left(\frac{1}{p_i p_j} -  \frac{1}{p_{ij}}  \right)m_im_j,
\]

where $p_{ij}$ is the probability to include both $i$ and $j$ in the sample, or in our case, to detect both trees $i$ and $j$. The indexing goes over the detected trees, not the whole population. Through conditional probability we can write $p_{ij} = p_{j|i}p_i$, where $p_{j|i}$ is the probability of detecting $j$ when $i$ is detected. Because of our sequential construction the probability of detecting $j$ always takes into account the fact that $i$ has been detected for $j>i$. Hence, $p_{j|i} = p_j$ and $p_{ij} = p_i p_j$, and furthermore

\begin{equation}
\label{eqn:variance}
\widehat{var}(\widehat{\tau}) = \sum_{i=1}^n \left(\frac{1}{p_i^2} -  \frac{1}{p_i}\right)m_i^2.
\end{equation}

The estimated variance makes it possible to produce approximate confidence intervals \citep[][p. 70]{Thompson2012}

\[
\widehat{\tau} \pm t_{\alpha_2, n-1} \sqrt{\widehat{var}(\widehat{\tau})},
\]

where $t_{\alpha_2, n-1}$ is the critical value corresponding to the selected nominal confidence level from the $t$-distribution with $n-1$ degrees of freedom. The $t$-distribution can be replaced with a standard normal distribution if the number of detected trees is large. We follow the pragmatic rule given by \citet{Thompson2012} and use the $t$-distribution if the number of detected trees is less than $50$. We note that a large portion of the simulated circular plots we use here for testing have less than $50$ detected trees. 

\subsection{Comparison to the estimator of Kuronen et al.}

The estimator of \citet{Kuronenetal2019} is of the same general form as ours (Equation~\ref{eqn: HTest}). However, instead of weighting the detections with detection probabilities $p_{\alpha}(r_i, d_i)$ which we have formulated in Equation~\ref{eqn: datadete}, the detections are weighted with weights $w_i$ that correspond to surface areas of the visible (or nonvisible) planar set that has been morphologically transformed according to the detection condition. Using the notation that we have presented the weights can be written as

\begin{equation}\label{eqn: kuroweights}
w_i   =  \left\lbrace\begin{array}{ll}
1-\frac{|(\cup_{j=1}^{N}A_j)\ominus B(o,  d_i/2)|}{|W|}, & \alpha=-1 \\ 
1-\frac{|\cup_{j=1}^{N}A_j|}{|W|}, & \alpha=0 \\ 
1-\frac{|(\cup_{j=1}^{N}A_j)\oplus B(o,  d_i/2)|}{|W|}, & \alpha=1
\end{array}\right.,
\end{equation}

corresponding to the detection conditions of any, centre and full visibility. \citet{Kuronenetal2019} did not include the parameter $\alpha$ to their construction, but considered the detection conditions as totally separate cases. They did, however, consider  a different kind of a weighting for the partial visibility detection condition, related to the visible proportion of the boundary of the stem disc, whereas in our construction partial visibility cases are handled by setting $\alpha$ to a value that differs from $-1$, $0$ and $1$ and are related to the proportion of visible stem disc radius.  

Comparison of our weights in Equation~\ref{eqn: datadete} and the weights of \citet{Kuronenetal2019} in Equation~\ref{eqn: kuroweights} shows that the main difference between the two constructions is the sequential nature of our detection probabilities: in our construction only the trees that are before tree $i$ in the ordering affect the weight of tree $i$, whereas in the construction of \citet{Kuronenetal2019} all of the trees, including tree $i$, have an effect. The premises of the constructions are also different. Whereas we approach the situation from a probabilistic perspective, \citet{Kuronenetal2019} are following the idea that a tree that is detected should belong to such a sampling window from which it can be detected, and hence, the size of the sampling window should change for every tree. \citet{Kuronenetal2019} was not able to show unbiasedness of their estimator -- in fact, they showed that the estimator has positive bias even in the case of Poisson process data -- and they did not present an estimator for the variance either.

\subsection{Characterizing deviations from the Poisson assumption}

As our estimator can be shown to be unbiased when the data is generated by a Poisson process, it is of interest to see how the bias of the estimator behaves when the spatial structure of the data deviates from this Poisson assumption to either clustering or regularity, and if the magnitude of this deviation has an effect on the magnitude of the bias. We use the $L$-function, derived from the $K$-function as
\[L(r)=\sqrt{\frac{K(r)}{\pi}}, \quad r \geq 0\]
to characterize the spatial structure of a point pattern in a plot \citep[see e.g.][Chapter 4.3]{Illianetal2008}. The $K$-function is related to the expected number of points of a point process that are closer than distance $r$ to a point of the process. The $L$-function of the Poisson process is known: $L(r) = r$. For a clustered process $L(r)>r$, and for a regular process $L(r)<r$.

We estimated the $L$-functions for all the simulated plots and point patterns by using the R package spatstat \citep{Baddeleyetal2015} with isotropic edge-corrections \citep[see e.g.][Section 4.2.2]{Illianetal2008} with $r$ ranging from $0$ to $5$ metres. If the point pattern contained only one point, a case where $L$-function can not be estimated, we took the convention of characterizing it as coming from a Poisson process and set $\widehat{L}(r)=r$. To produce measures of deviation from the Poisson process, we used signed maximum deviation between $\widehat{L}$ and the identity line. First find the distance at which maximum absolute deviation occurs,
\[r^\ast = \arg \max_{r\in[0, 5]} |r - \hat{L}(r) |.\]
Then, the $L$-based measure of deviation is given by
\begin{equation}
\label{eqn: Lmeasure}
r^\ast - \hat{L}(r^\ast),
\end{equation}
producing positive values if the pattern shows stronger signs of regularity than clustering, and negative values if the pattern shows stronger signs of clustering than regularity.

We also used this deviation measure to characterize whole data sets by first calculating the mean $L$-function over all $n$ plots of the data as
\[\bar{\hat{L}}(r) = \frac{1}{n}\sum_{i=1}^{n} \hat{L}_i(r)\]
and then calculating the deviation measure for the mean $L$-function. The mean $L$-function can be seen as an estimator of the expected $L$-function of the process that has generated the data, and as $L(r)=r$ for Poisson process is also an expectation, this is quite natural way to characterize the deviation of a process from the Poisson process. The $L$-functions were estimated in R \citep{R2019} with the function \textbf{Lest} in the R package \textbf{spatstat} \citep{Baddeleyetal2015}.

\section{Materials}

\subsection{Field data}

We use field data measured in 2017 in Eastern Finland. It consists of 111 square $30 \times 30$ metre plots placed to the area of about 43,000 hectares. The plots were sampled from a systematic network using a priori information of stand maturity and dominant tree species.  Tree species were determined and tree height and DBH were measured for all trees with a DBH $\geq$ 5 cm. Tree locations were determined using a variant of the triangulation method proposed by \citet{Korpelaetal2007}. Field measurements are documented in detail in \citet{Ratyetal2019}.

The central $10 \times 10$ metre square of every plot was covered with a triangular grid where the distance between points was at least 1 metre. Every grid point  was used as a centre for a circular sampling plot with radius 10 metres. If a tree stem covered the centre point, it was removed. This produced $111 \times 126 = 13 986$ simulated plots using true field data. Statistics related to $N$ and $G$ in these plots are presented in Table~\ref{table:sup1}.

\subsection{Process simulated data}

Circular field plots of radius $10$ metres were simulated from four different types of marked point processes: Poisson process, nonoverlapping discs process, Gibbs hard core process and Log-Gaussian Cox process (generation of point patterns from these processes is more thoroughly depicted in Section~\ref{sec:gmpp}). Two different variants of the Gibbs hard core process and five different variants of the Log-Gaussian Cox process were simulated. In all cases, process intensities ($N$) $500$, $1000$, $1500$, $2000$, $2500$, $3000$, $3500$, $4000$, $4500$, and $5000$ $\mathrm{stems} \cdot \mathrm{ha}^{-1}$ were used. These intensities were chosen to cover the variation in stem densities in natural forests and have been previously used by e.g. \citet{Olofssonandolsson2018}. The number of plots generated was $10000$ for the Poisson process, $1000$ with every intensity, and $2000$ per other point processes, $200$ with every intensity. Special attention was given for the Poisson process because it is theoretically in line with the assumptions of our estimator and hence demonstrating the performance of the estimator in Poisson process data is of great importance. In all cases the point process was simulated in a larger simulation window to take into account possible effects at the plot boundary. The trees that did not come into contact with the plot boundary at $10$ metres, and hence could not have any effect on the estimation, were removed. The number of points in the simulation window was always first generated from a Poisson distribution with expected value $N|W|$, where $|W|$ is the area of the simulation window. Only point patterns where the centre point of the plot was not covered by a stem disc were accepted. All simulations were done with R \citep{R2019}. Statistics related to the simulated $N$ and $G$ are presented in Table~\ref{table:sup2} and example plots are presented in Figure~\ref{fig:sup1}.

\subsubsection{Generating tree diameters}

An adjusted version of the parameter recovery method of \citet{Siipilehtoetal2013} was followed to produce Weibull distributions for the DBH. Mean DBH ($D$) and basal area ($G$) pairs of $6$ cm and $3$ $m^2 \cdot \mathrm{ha}^{-1}$, $12$ cm and $12$ $m^2 \cdot \mathrm{ha}^{-1}$, $15$ cm and $20$ $m^2 \cdot \mathrm{ha}^{-1}$, and $21$ cm and $35$ $m^2 \cdot \mathrm{ha}^{-1}$ were taken as a starting point. Then, for shape parameter $\gamma$ and scale parameter $\beta$ the sum of differences between the means and quadratic means
\[|\beta \Gamma(1+1/\gamma) - D)|+|\beta^2\Gamma(1+2/\gamma) - G/(N \pi/40000)|\] 
was minimized. This produced $40$ different DBH distributions with expected basal areas almost exactly agreeing with the four given mean values of $G$ and expected DBH differing from the given mean values of DBH  to keep $N$ and $G$ compatible. In other words, for every intensity four different DBH distributions corresponding to different mean DBH and basal area were used to simulate the DBH marks for the trees. It should be noted that these intensities and distributions are parameters of the random processes and hence the simulated stem densities, mean DBH and basal areas will be different from these numbers and vary according to the nature of the process.

\subsubsection{\label{sec:gmpp} Generating marked point patterns}

\textit{Poisson process} (abbreviated as \textit{Poisson} in the tables and figures) forests with complete spatial randomness of points and independent and identically distributed DBH were generated in a circular window of radius $11$ metres, with the given intensities and DBH distributions. \textit{Nonoverlapping discs process} (\textit{Nonoverlapping}), producing point patterns where the stem discs can not overlap, was used to simulate forests with a more realistic hard core property. The first point location was generated uniformly in a circular window of radius $11$ metres and mark was generated from the DBH distribution. For points $2,\ldots, M$ a suggested point with a mark from the corresponding DBH distribution was generated uniformly in the same window. If the point was located in such a way that its stem disc did not overlap the previous discs it was accepted, otherwise it was rejected and the marked point was simulated again. At every step  $2,\ldots, M$ $10000$ attempts to simulate were allowed. If the algorithm failed to simulate an acceptable point during these attempts the simulation was regarded as finished.

Another type of hard core process where the points are at least $1$ metre apart was simulated as a \textit{Gibbs hard core process} (\textit{Gibbs 1}). The points were first simulated in a $40\times 40$ metre square window. The locations of points were simulated from a Gibbs process with known number of points \citep[see e.g.][]{Illianetal2008}.  Then an independent sample was taken from the DBH distribution to assign marks to the points. Similar scheme was used to simulate a Gibbs $1.5$ metre hard core process (\textit{Gibbs 1.5}).

\textit{Log-Gaussian Cox process} (\textit{Cluster}) with Matern covariance function was used to produce plots with spatial clustering. The smoothness and variance parameters of the covariance function were fixed to $2$ and $1$, whereas the range was varied between $2$, $4$, $6$, $8$, and $10$ metres, producing five different variants of a cluster process.  The points were simulated in a $40\times 40$ metre square window. The Gaussian field was simulated with the R package \textbf{RandomFields} \citep{RandomFields2019} to produce the density field over the window. The points were then generated according to that density with the \textbf{spatstat} function \textbf{rpoint} \citep{Baddeleyetal2015}.

\newpage
\section{Evaluation of results}

Relative root-mean-squared errors

\[
RMSE\% = \frac{100}{\bar{y}}\times\sqrt{\frac{\sum_{i=1}^n(\hat{y}_i - y_i)^2}{n}}
\]
 
and relative means of errors

\[
ME\% = \frac{100}{\bar{y}}\times\frac{\sum_{i=1}^n(\hat{y}_i - y_i)}{n}
\]

where $\hat{y}_i$ is the estimated value, $y_i$ the true value, $\bar{y}$ the mean of the true values and $n$ the number of plots, were used to evaluate and compare the estimation errors between different estimators. The estimators that we benchmark our method against are the estimators of \citet{Olofssonandolsson2018} and \citet{Kuronenetal2019}. These estimators are the same when the detection condition is centre point visibility. We also report the errors of estimation based on the detected trees with no corrections. It should be noted that both \citet{Olofssonandolsson2018} and \citet{Kuronenetal2019} assumed that the stem discs belong to the visible area, whereas we assume that they belong to the nonvisible area. To compare only the differences of distance-based and area-based construction on the estimation we wanted to keep all other effects the same and decided to modify these estimators to include the stem discs to the nonvisible area. This should not affect the results greatly; the important part is to have the detection conditions matching the nonvisible areas, in other words that the simulated trees have been correctly classified as either detected or not detected. In the result tables we refer to our estimator as ''HT'', the estimator of \citet{Olofssonandolsson2018} as ''O \& O'', the estimator of  \citet{Kuronenetal2019} as ''Kuronen et al.'', and finally the estimator based on just the detected trees as ''detected''. We evaluate the performance of the estimators with detection conditions full, centre, and any. The estimators of \citet{Olofssonandolsson2018} and \citet{Kuronenetal2019} are the same when the detection condition is centre point visibility.

\section{Results}
In the plots based on the field measured data, the HT-like estimator produces better or very similar error values in most cases (Table~\ref{table:liperires}). The exception is the estimation of $G$ when the detection condition is full visibility, where the estimator of \citet{Olofssonandolsson2018} produces lowest error values. 
 
\begin{table}[!h]
\centering
\caption{\label{table:liperires} Estimation errors over the simulated plots based on the field data in Liperi for detection conditions \emph{full} visibility, \emph{centre} point visibility and \emph{any} visibility, and for estimators by us (\emph{HT}), \citet{Olofssonandolsson2018} (\emph{O \& O}), and \citet{Kuronenetal2019}. Lowest (absolute) values for each case are bolded. The estimators O \& O and Kuronen et al. coincide in the centre point case.}

\begin{tabular}{lll rrrr}
  \hline
        &                &  & \multicolumn{2}{c}{N}  & \multicolumn{2}{c}{G}   \\ [1pt]
        \cline{4-5} \cline{6-7} \\ 
  Condition   & Estimator      &           &  RMSE\% &  ME\% & RMSE\% &  ME\% \\ 
   \hline
full   & HT             &           &   \textbf{8.0} &   \textbf{2.2} &  11.0 &   4.1 \\ 
         & O \& O        &           &   9.0 &  -3.7 &   \textbf{9.6} &  \textbf{-2.7} \\ 
         & Kuronen et al. &           &   8.5 &   3.3 &  12.0 &   5.8 \\ 
         & detected        &           &  22.5 & -15.4 &  18.3 & -13.5 \\ 
  centre & HT             &           &   \textbf{6.2} &   \textbf{1.2} &   \textbf{7.6} &   \textbf{2.1} \\  
         & Kuronen et al. &           &   6.3 &   1.6 &   7.8 &   2.8 \\ 
         & detected           &           &  16.4 & -10.8 &  12.6 &  -8.7 \\ 
  any    & HT             &           &   \textbf{4.6} &   \textbf{0.3} &   \textbf{4.8} &   \textbf{0.4} \\ 
         & O \& O       &           &  10.2 &   6.5 &  10.4 &   7.4 \\ 
         & Kuronen et al. &           &   4.6 &   0.4 &   4.8 &   0.6 \\ 
         & detected           &           &  10.7 &  -6.7 &   7.7 &  -4.7 \\ 
   \hline
\end{tabular}
\end{table}

The results in the plots simulated from marked point processes are presented in Table~\ref{table:poissonres} for the Poisson process data and Tables \ref{table:sup3}, \ref{table:sup4}, and \ref{table:sup5},  for the others. When comparing the HT-like estimator and the estimator of \citet{Kuronenetal2019}, the HT-like estimator produces better or very similar error values in data simulated from the Poisson process, Gibbs processes and the nonoverlapping discs model. Especially in the Poisson process case the ME\% values of stem density estimation are very close to zero, which is consistent with the unbiasedness of the estimator for homogeneous Poisson process. In the clustered data, the estimator of \citet{Kuronenetal2019} produces lower error values than the HT-like estimator. Both of the estimators produce larger estimation errors in the clustered data than in the other types of data. Both of the estimators show overestimation in the regular data and underestimation in the clustered data. The estimator of \citet{Olofssonandolsson2018} does not show this behaviour, but shows overestimation with full visibility detection condition and underestimation with any visibility detection condition in all of the simulated data sets. For the Gibbs hard core data and full visibility detection condition it produces lower error values than the other two estimators, except for ME\% of $G$.

\begin{table}
\centering
\caption{\label{table:poissonres} Estimation errors over the Poisson process simulated plots  for detection conditions \emph{full} visibility, \emph{centre} point visibility and \emph{any} visibility, and for estimators by us (\emph{HT}), \citet{Olofssonandolsson2018} (\emph{O \& O}), and \citet{Kuronenetal2019}. Lowest (absolute) values for each case are bolded. The estimators O \& O and Kuronen et al. coincide in the centre point case.}
\begin{tabular}{lllrrrr}
  \hline
                         &                &  & \multicolumn{2}{c}{N}  & \multicolumn{2}{c}{G} \\ 
 \cline{4-5} \cline{6-7} \\
 Condition   & Estimator      &           & RMSE\% & ME\% & RMSE\% & ME\% \\ 
   \hline
full   & HT             &           &   \textbf{6.1} &   \textbf{0.0} &  \textbf{13.6} &   \textbf{0.3} \\ 
       & O \& O        &           &  11.1 &  -7.5 &  16.1 &  -8.3 \\ 
       & Kuronen et al. &           &   6.1 &   0.7 &  16.5 &   1.7 \\ 
       & detected           &           &  28.3 & -21.5 &  34.0 & -23.3 \\ 
centre & HT             &           &   4.8 &   \textbf{0.0} &   \textbf{7.8} &   \textbf{0.1} \\
       & Kuronen et al. &           &   \textbf{4.8} &   0.2 &   8.2 &   0.6 \\ 
       & detected           &           &  20.3 & -15.1 &  23.8 & -16.0 \\ 
any    & HT             &           &   3.4 &   \textbf{0.0} &   \textbf{5.0} &   \textbf{0.0} \\ 
       & O \& O        &           &  12.0 &   8.5 &  15.2 &   9.6 \\ 
       & Kuronen et al. &           &   \textbf{3.4} &   0.1 &   5.1 &   0.2 \\ 
       & detected           &           &  11.9 &  -8.4 &  14.1 &  -8.8 \\ 
  \hline
\end{tabular}
\end{table}

\begin{figure}
\centerline{\includegraphics[width=14cm]{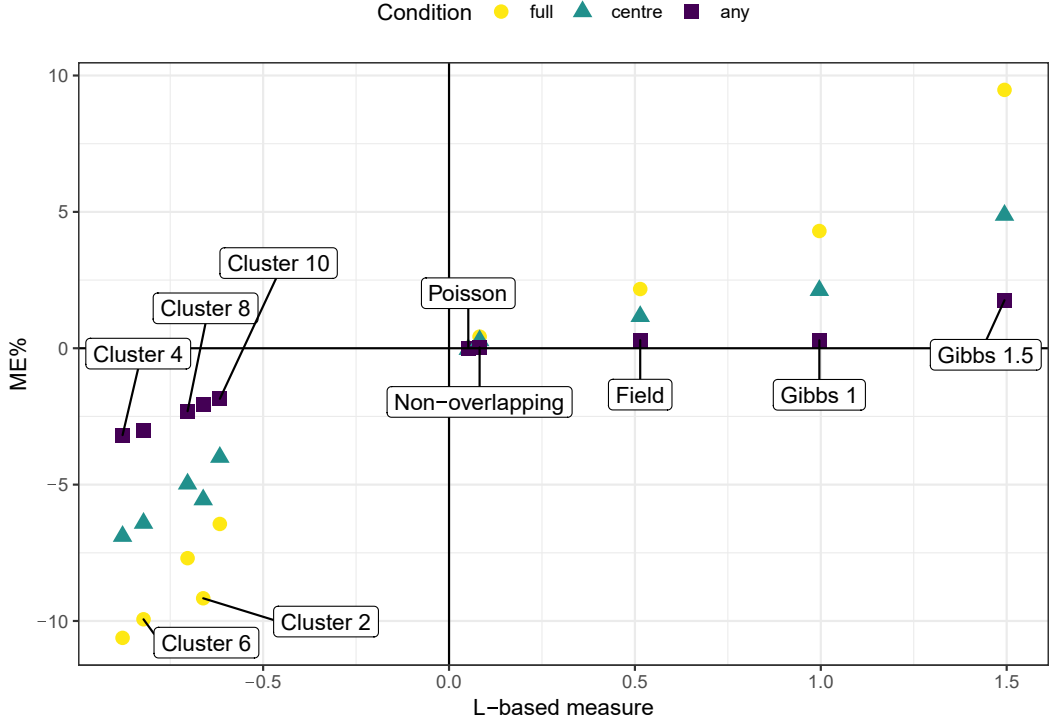}}
\caption{\label{fig:LvsME} The relationship between the $L$-function based deviation measure (Equation~\ref{eqn: Lmeasure}), describing the deviation of a data set from Poisson process to regularity or clustering, and relative mean of errors of the Horvitz--Thompson-like estimates for detection conditions \emph{full} visibility, \emph{centre} point visibility and \emph{any} visibility.}
\end{figure}

Figure~\ref{fig:LvsME} represents the relationship between bias and deviation of a data set from the Poisson process assumption. The ME\% of the different data sets are plotted against the $L$-function based deviation measures calculated from the mean $L$-functions of the data sets. The Poisson process data, having deviation measure closest to zero, has ME\% closest to zero, indicating unbiasedness. When the deviation measure grows larger, the errors start to also grow, showing that larger deviations to regularity produce larger overestimations. On the other hand, deviations to clustering produce underestimation, and the magnitude of deviation affects the magnitude of underestimation. The deviation and detection condition also have a connection, the any visibility detection condition results being least affected by the magnitude of clustering or regularity, and full visibility detection condition being affected the most. The field data based data set is somewhat regular, having $L$-function based deviation of roughly $0.5$. It should be noted that although the data sets have ''on average'' a certain type of behaviour, e.g. heavily clustered or slightly regular, the plots in these data sets can have widely varying deviation measures (see Tables \ref{table:sup1} and \ref{table:sup2}). However, the mean behaviour seems to have the largest effect on the bias of the estimator.

To examine the estimation of estimator variance and the goodness of the approximate confidence intervals, we calculated the $90$, $95$ and $99$ per cent intervals for stem density and basal area in the simulated data sets with different visibility conditions and then calculated how many of the simulated stem densities and basal areas belonged to their corresponding confidence intervals. The results about stem density are presented in Table~\ref{table:intervals} and Table~\ref{table:sup6}. In the Poisson and nonoverlapping discs data sets the observed coverage probabilities of the confidence intervals are very close to the nominal confidence levels with all of the detection conditions tested.  In the slightly regular field data set the percentages are mostly in line with the $99$ per cent confidence level, but the $90$ and $95$ per cent intervals are slightly conservative. In the Gibbs hard core process data sets with strong regularity the visibility condition has a larger effect on the goodness of the interval estimates. This can be especially seen in the $90$ per cent interval case, where under the full visibility condition only $87.1$ per cent of the simulated stem densities belong to their corresponding intervals, whereas under the any visibility condition the value is $93.2$ per cent, going from anti-conservative to conservative intervals. However, the $99$ per cent intervals for Gibbs $1$ metre hard core process have very similar nominal and observed coverage probabilities. For Gibbs $1.5$ metre hard core process, most of the intervals are anti-conservative.  In the clustered data sets, all of the intervals are anti-conservative, indicating that the variances have been underestimated in addition to the underestimation present in the estimated stem densities. The results relating to basal area estimation, presented in Table~\ref{table:sup7}, show similar effects than the stem density estimation results.

\begin{table}
\centering
\caption{\label{table:intervals} The percentages of simulated stem densities belonging to their approximate confidence intervals for detection conditions \emph{full} visibility, \emph{centre} point visibility and \emph{any} visibility.}

\begin{tabular}{lll rrr}
  \hline
 &  & & \multicolumn{3}{c}{interval}\\
\cline{4-6} \\
Data                 & Condition       &  & 90\% & 95\% & 99\% \\ 
   \hline
Field          & full   &          & 94.0 & 97.5 & 99.5 \\ 
                  & centre &          & 94.1 & 97.4 & 99.3  \\ 
                  & any    &          & 93.4 & 96.5 & 98.8 \\ 
Poisson         & full   &       & 90.0 & 94.9 & 98.7 \\ 
                  & centre &       & 89.9 & 94.5 & 98.7 \\ 
                  & any    &       & 90.5 & 94.9 & 98.4 \\  
   \hline
\end{tabular}
\end{table}

\begin{figure}
\centerline{\includegraphics[width=14cm]{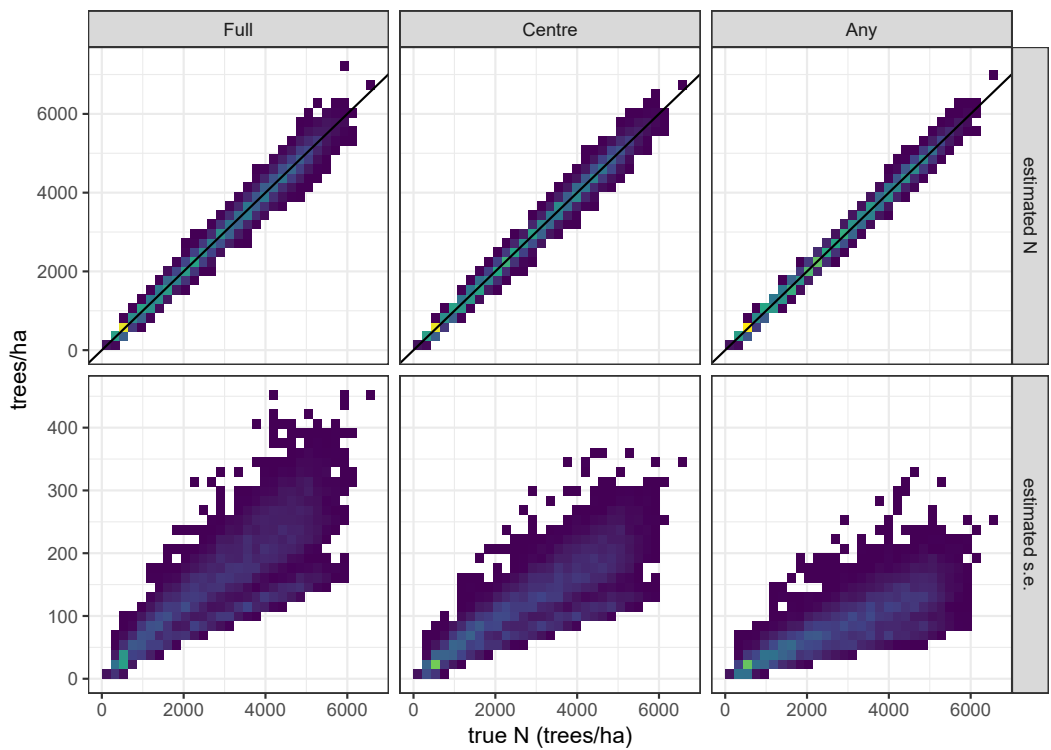}}
\caption{\label{fig:poissonfigures} Horvitz--Thompson-like estimates of stem densities and their estimated standard errors as functions of the simulated stem density in the Poisson process data for detection conditions \emph{full} visibility, \emph{centre} point visibility and \emph{any} visibility. Areas that are lighter contain more sample plots than the darker areas.}
\end{figure}

The relationships of the estimated stem densities and their estimated standard errors to the simulated stem densities in the Poisson process data are presented in Figure~\ref{fig:poissonfigures}. The estimated stand densities are well centred around the identity line, with more variance in the estimates as the simulated stem density increases. The estimation results get better when going from full visibility to centre point visibility, and further to any visibility detection condition, represented by the more concentrated point cloud. This effect is also shown by the incrementally smaller RMSE\% values (Table~\ref{table:poissonres}). The estimated standard errors increase as the simulated stand density increases, as does the variation in the estimated standard errors. The errors decrease when moving from the full visibility to any visibility detection condition.

\section{Discussion}

Our estimator has a clear connection to distance sampling \citep[see e.g.][]{Bucklandetal2004}, which is commonly used in estimating for example sizes of wild animal populations from a sample of observations at different distances from the observer. Most commonly a parametric model describing the probability to detect an animal at a certain distance is fitted to the sample data, and then a Horvitz--Thompson-like estimator is used, either with the fitted probabilities for each observation depending on its distance, or the mean probability for every observation. Fitting of a parametric detection probability model is also the approach of \citet{Duncansonetal2014} and \citet{Astrupetal2014} to estimating population totals from TLS data in forestry. Our construction could be seen as a nonparametric distance sampling estimator - provided that the value of the tuning parameter is known - where the detection probabilities are calculated directly from the data for each observation individually. It could also be seen as distance sampling with empirical detection probabilities under an assumed Poisson model.

The strength of our construction, when compared to \citet{Kuronenetal2019} and other previously published methods, is the ability to prove the unbiasedness of the estimator when the data is generated by a Poisson process -- \citet{Kuronenetal2019} showed that there is a positive bias in their estimator even in this case -- and a way to calculate approximate confidence intervals. In both cases the sequential nature of the detection probabilities, i.e. only the trees closer to the plot centre affecting the detection of trees further away, is of great use. In addition to the unbiasedness in Poisson process data, Figure~\ref{fig:LvsME} shows that there is a natural relation between deviations from Poisson process and the bias of the estimator: regularity leads to overestimation, clustering to underestimation.

The HT-like estimator with distance-based detection probabilities produces in most simulation cases estimation errors that are lower than or comparable to the errors produced by  the estimators of \citet{Olofssonandolsson2018} and \citet{Kuronenetal2019}. The estimator works very well in the Poisson process data, as is to be expected. The estimator is slightly biased in the slightly regular field data. However, the bias is to a ''safe'' direction because the confidence intervals are conservative in this case. The results indicate that the estimator should not be used ''as is'' for strongly clustered or regular data. In these cases, it might be possible to \emph{tune} the estimator by adjusting the parameter $\alpha$. An example of unintentional tuning is for example the estimator of \citet{Olofssonandolsson2018} producing lower ME\% of $G$ in the field data with full visibility detection condition. This is because the correction is ''wrong'': when compared to the area from which trees can be detected based on the detection condition a larger area is used for weighting, pushing the estimates down. Intentional tuning in our case would entail calculating estimates with several values of parameter $\alpha$ in a training data set and choosing a value that minimizes some error value as the ''right'' detection condition for that type of data.

Figure~\ref{fig:poissonfigures} further demonstrates the properties of our estimator when the data is Poisson distributed. The unbiasedness is shown by the concentration of the stem density estimates along the identity lines. The increasing simulated stem density produces increasing estimates of standard error. As the number of stems increases, so does the number of detected stems, which introduces more positive terms to the formula of the variance estimator (Equation~\ref{eqn:variance}), which produces a larger variance. The increase caused by a single tree is connected to its detection probability; small probability leads to a large increase.

The construction could be generalized in several ways. For example, other objects than tree stems producing nonvisible areas behind them could be added in when calculating the detection probabilities, as long as the object is detected from the data in such a way that the geometry of the nonvisible area it produces is known. The visibility condition of a tree could depend on its size and distance from the scanner. Other effects connecting distance and detection probability, such as the number of laser returns per unit area which diminishes as a function of distance, could be added to the calculation of detection probabilities, either through the parameter $\alpha$ or as additional weights.

\section{Conclusions}
We have presented a Horvitz--Thompson-like estimator with distance-based detection probabilities for single scan terrestrial laser scanning and evaluated its performance in a simulation study consisting of plots based on field measured data and simulated from several different point processes. The estimator produces error values comparable to or lower than the two benchmark methods in most cases. A notable exception is the clustered data. The estimator is unbiased when the data comes from a Poisson process. The variance of the estimator can be estimated and approximate confidence intervals can be formed.  The detection probability construction is easily generalizable to allow objects other than tree stems to affect detection and addition of different distance-based effects on detection. The estimator has a tuning parameter which could be used to find an appropriate detection condition for a scanner in training data.

\section*{Acknowledgements}
This work was funded by the Academy of Finland (decision number 310073). We would like to thank Juha Heikkinen, Mikko Kuronen and Mari Myllymäki for useful discussions on the topic.

\bibliographystyle{biom} 
 \bibliography{mybib}

\begin{thebibliography}{}

\bibitem[\protect\citeauthoryear{Astrup, Ducey, Granhus, Ritter, and von
  Lüpke}{Astrup et~al.}{2014}]{Astrupetal2014}
Astrup, R., Ducey, M.~J., Granhus, A., Ritter, T., and von Lüpke, N. (2014).
\newblock Approaches for estimating stand-level volume using terrestrial laser
  scanning in a single-scan mode.
\newblock {\em Canadian Journal of Forest Research} {\bf 44,} 666--676.

\bibitem[\protect\citeauthoryear{Baddeley, Rubak, and Turner}{Baddeley
  et~al.}{2015}]{Baddeleyetal2015}
Baddeley, A., Rubak, E., and Turner, R. (2015).
\newblock {\em Spatial Point Patterns: Methodology and Applications with {R}}.
\newblock Chapman and Hall/CRC Press, London.

\bibitem[\protect\citeauthoryear{Buckland, Anderson, Burnham, Laake, Borchers,
  and Thomas}{Buckland et~al.}{2004}]{Bucklandetal2004}
Buckland, S.~T., Anderson, D.~R., Burnham, K.~P., Laake, J.~L., Borchers,
  D.~L., and Thomas, L. (2004).
\newblock {\em Advanced distance sampling}.
\newblock Oxford University Press.

\bibitem[\protect\citeauthoryear{Duncanson, Cook, Hurtt, and Dubayah}{Duncanson
  et~al.}{2014}]{Duncansonetal2014}
Duncanson, L., Cook, B., Hurtt, G., and Dubayah, R. (2014).
\newblock An efficient, multi-layered crown delineation algorithm for mapping
  individual tree structure across multiple ecosystems.
\newblock {\em Remote Sensing of Environment} {\bf 154,} 378 -- 386.

\bibitem[\protect\citeauthoryear{Gregoire and Valentine}{Gregoire and
  Valentine}{2007}]{GregoireandValentine2007}
Gregoire, T. and Valentine, H. (2007).
\newblock {\em Sampling Strategies for Natural Resources and the Environment}.
\newblock Chapman and Hall/CRC, New York.

\bibitem[\protect\citeauthoryear{Illian, Penttinen, Stoyan, and Stoyan}{Illian
  et~al.}{2008}]{Illianetal2008}
Illian, J., Penttinen, A., Stoyan, H., and Stoyan, D. (2008).
\newblock {\em Statistical Analysis and Modelling of Spatial Point Patterns}.
\newblock John Wiley \& Sons, Ltd.

\bibitem[\protect\citeauthoryear{Kansanen, Vauhkonen, L\"ahivaara, and
  Meht\"atalo}{Kansanen et~al.}{2016}]{Kansanenetal2016}
Kansanen, K., Vauhkonen, J., L\"ahivaara, T., and Meht\"atalo, L. (2016).
\newblock Stand density estimators based on individual tree detection and
  stochastic geometry.
\newblock {\em Canadian Journal of Forest Research} {\bf 46,} 1359--1366.

\bibitem[\protect\citeauthoryear{Korpela, Tuomola, and Välimäki}{Korpela
  et~al.}{2007}]{Korpelaetal2007}
Korpela, I., Tuomola, T., and Välimäki, E. (2007).
\newblock Mapping forest plots: An efficient method combining photogrammetry
  and field triangulation.
\newblock {\em Silva Fennica} {\bf 41,}.

\bibitem[\protect\citeauthoryear{Kuronen, Henttonen, and Myllymäki}{Kuronen
  et~al.}{2019}]{Kuronenetal2019}
Kuronen, M., Henttonen, H.~M., and Myllymäki, M. (2019).
\newblock Correcting for nondetection in estimating forest characteristics from
  single-scan terrestrial laser measurements.
\newblock {\em Canadian Journal of Forest Research} {\bf 49,} 96--103.

\bibitem[\protect\citeauthoryear{{Liang}, {Litkey}, {Hyyppa}, {Kaartinen},
  {Vastaranta}, and {Holopainen}}{{Liang} et~al.}{2012}]{Liangetal2012}
{Liang}, X., {Litkey}, P., {Hyyppa}, J., {Kaartinen}, H., {Vastaranta}, M., and
  {Holopainen}, M. (2012).
\newblock Automatic stem mapping using single-scan terrestrial laser scanning.
\newblock {\em IEEE Transactions on Geoscience and Remote Sensing} {\bf 50,}
  661--670.

\bibitem[\protect\citeauthoryear{Lovell, Jupp, Newnham, and Culvenor}{Lovell
  et~al.}{2011}]{Lovelletal2011}
Lovell, J., Jupp, D., Newnham, G., and Culvenor, D. (2011).
\newblock Measuring tree stem diameters using intensity profiles from
  ground-based scanning lidar from a fixed viewpoint.
\newblock {\em ISPRS Journal of Photogrammetry and Remote Sensing} {\bf 66,} 46
  -- 55.

\bibitem[\protect\citeauthoryear{Olofsson and Olsson}{Olofsson and
  Olsson}{2018}]{Olofssonandolsson2018}
Olofsson, K. and Olsson, H. (2018).
\newblock Estimating tree stem density and diameter distribution in single-scan
  terrestrial laser measurements of field plots: a simulation study.
\newblock {\em Scandinavian Journal of Forest Research} {\bf 33,} 365--377.

\bibitem[\protect\citeauthoryear{{R Core Team}}{{R Core Team}}{2019}]{R2019}
{R Core Team} (2019).
\newblock {\em R: A Language and Environment for Statistical Computing}.
\newblock R Foundation for Statistical Computing, Vienna, Austria.

\bibitem[\protect\citeauthoryear{Raumonen, Casella, Calders, Murphy,
  {\AA}kerblom, and Kaasalainen}{Raumonen et~al.}{2015}]{Raumonenetal2015}
Raumonen, P., Casella, E., Calders, K., Murphy, S., {\AA}kerblom, M., and
  Kaasalainen, M. (2015).
\newblock Massive-scale tree modelling from tls data.
\newblock {\em ISPRS Annals of the Photogrammetry, Remote Sensing and Spatial
  Information Sciences} {\bf II 3,} 189--196.

\bibitem[\protect\citeauthoryear{Ravaglia, Bac, and Fournier}{Ravaglia
  et~al.}{2017}]{Ravagliaetal2017}
Ravaglia, J., Bac, A., and Fournier, R.~A. (2017).
\newblock Extraction of tubular shapes from dense point clouds and application
  to tree reconstruction from laser scanned data.
\newblock {\em Computers \& Graphics} {\bf 66,} 23 -- 33.
\newblock Shape Modeling International 2017.

\bibitem[\protect\citeauthoryear{Räty, Packalen, and Maltamo}{Räty
  et~al.}{2019}]{Ratyetal2019}
Räty, J., Packalen, P., and Maltamo, M. (2019).
\newblock Nearest neighbor imputation of logwood volumes using bi-temporal als,
  multispectral als and aerial images.
\newblock {\em Scandinavian Journal of Forest Research} {\bf 34,} 469--483.

\bibitem[\protect\citeauthoryear{Schlather, Malinowski, Oesting, Boecker,
  Strokorb, Engelke, Martini, Ballani, Moreva, Auel, Menck, Gross, Ober,
  Ribeiro, Ripley, Singleton, Pfaff, and {R Core Team}}{Schlather
  et~al.}{2019}]{RandomFields2019}
Schlather, M., Malinowski, A., Oesting, M., Boecker, D., Strokorb, K., Engelke,
  S., Martini, J., Ballani, F., Moreva, O., Auel, J., Menck, P.~J., Gross, S.,
  Ober, U., Ribeiro, P., Ripley, B.~D., Singleton, R., Pfaff, B., and {R Core
  Team} (2019).
\newblock {\em {RandomFields}: Simulation and Analysis of Random Fields}.
\newblock R package version 3.3.6.

\bibitem[\protect\citeauthoryear{Seidel and Ammer}{Seidel and
  Ammer}{2014}]{Seidelandammer2014}
Seidel, D. and Ammer, C. (2014).
\newblock Efficient measurements of basal area in short rotation forests based
  on terrestrial laser scanning under special consideration of shadowing.
\newblock {\em iForest - Biogeosciences and Forestry} {\bf 7,} 227--232.

\bibitem[\protect\citeauthoryear{Siipilehto and Meht\"atalo}{Siipilehto and
  Meht\"atalo}{2013}]{Siipilehtoetal2013}
Siipilehto, J. and Meht\"atalo, L. (2013).
\newblock Parameter recovery vs. parameter prediction for the weibull
  distribution validated for {S}cots pine stands in {F}inland.
\newblock {\em Silva Fennica} {\bf 47,}.

\bibitem[\protect\citeauthoryear{Strahler, Jupp, Woodcock, Schaaf, Yao, Zhao,
  Yang, Lovell, Culvenor, Newnham, Ni-Miester, and Boykin-Morris}{Strahler
  et~al.}{2008}]{Strahleretal2008}
Strahler, A.~H., Jupp, D.~L., Woodcock, C.~E., Schaaf, C.~B., Yao, T., Zhao,
  F., Yang, X., Lovell, J., Culvenor, D., Newnham, G., Ni-Miester, W., and
  Boykin-Morris, W. (2008).
\newblock Retrieval of forest structural parameters using a ground-based lidar
  instrument (echidna®).
\newblock {\em Canadian Journal of Remote Sensing} {\bf 34,} S426--S440.

\bibitem[\protect\citeauthoryear{Thompson}{Thompson}{2012}]{Thompson2012}
Thompson, S.~K. (2012).
\newblock {\em Sampling}.
\newblock Wiley, Hoboken, New Jersey, 3rd edition.

\end{thebibliography}
 
\section*{Supporting Information}

The functions related to our estimator are available in the R package \textbf{lmfor}, available on CRAN. See the documentation for function \textbf{HTest\_cps} for further information. \vspace*{-8pt}

\newpage
 
\appendix

\renewcommand\thefigure{A.\arabic{figure}}
\renewcommand\thetable{A.\arabic{table}}
\section*{Appendix: Additional tables and figure}
\setcounter{figure}{0}
\setcounter{table}{0} 
\begin{table}[!h]
\caption{\label{table:sup1}Statistics of stem density $N$, basal area $G$ and the $L$-function based deviation measure (Equation~(5)) in the plots derived from the field data.}
\centering
\begin{tabular}{lllll}
\hline
   & \textbf{mean} & \textbf{sd} & \textbf{min} & \textbf{max} \\ \hline
 $N$, $\mathrm{stems} \cdot \mathrm{ha}^{-1}$ & 1104.1  & 648.3 & 127.3 & 4360.8 \\
$G$, $m^2 \cdot \mathrm{ha}^{-1}$ & 24.3 & 8.5 & 3.3 & 54.9 \\
$L$ & 0.85 & 0.95 & -2.08 & 5.00 \\
\hline
\end{tabular}
\end{table}

\begin{table}
\caption{\label{table:sup2}Statistics of stem density $N$, basal area $G$ and the $L$-function based deviation measure (Equation~(5)) in the process simulated plots.}
\centering
\begin{tabular}{lllllll}
\hline
 & &  & \textbf{mean} & \textbf{sd} & \textbf{min} & \textbf{max} \\ \hline
Poisson         & N &  &  2745.6 & 1470.4 & 31.8 & 6493.5 \\ 
                  & G &  &    17.5 & 12.8 & 0.1  & 124.1 \\ 
                  & L &  &    0.1  & 0.5  & -3.2 & 5.0 \\ 
  Non-overlapping & N &  &  2753.5 &  1459.2 &   222.8 &  6525.4 \\ 
                  & G &  &    17.7 &    13.1 &     0.6 &   142.8 \\ 
                  & L &  &     0.2 &     0.5 &    -2.8 &     3.1 \\ 
  Gibbs 1        & N &  &  2695.7 &  1409.3 &   254.6 &  5443.1 \\ 
                  & G &  &    17.1 &    12.0 &     0.4 &    70.9 \\ 
                  & L &  &     1.0 &     0.2 &    -2.3 &     2.6 \\ 
  Gibbs 1.5      & N &  &  2677.9 &  1439.5 &   222.8 &  5220.3 \\ 
                  & G &  &    17.0 &    12.3 &     0.4 &    98.7 \\ 
                  & L &  &     1.7 &     0.3 &     1.5 &     3.7 \\ 
  Cluster 2       & N &  &  2786.3 &  1695.5 &   127.3 & 10663.4 \\ 
                  & G &  &    17.7 &    13.8 &     0.5 &    95.7 \\ 
                  & L &  &    -0.8 &     0.6 &    -4.2 &     4.8 \\ 
  Cluster 4       & N &  &  2715.3 &  1834.8 &    95.5 & 12732.4 \\ 
                  & G &  &    17.6 &    15.3 &     0.1 &   115.5 \\ 
                  & L &  &    -1.0 &     0.8 &    -4.4 &     5.0 \\ 
  Cluster 6       & N &  &  2693.7 &  2012.0 &    63.7 & 16902.3 \\ 
                  & G &  &    17.0 &    15.3 &     0.0 &    96.4 \\ 
                  & L &  &    -0.9 &     0.9 &    -8.4 &     5.0 \\ 
  Cluster 8       & N &  &  2618.7 &  1987.3 &    63.7 & 13114.4 \\ 
                  & G &  &    16.6 &    15.2 &     0.0 &   104.5 \\ 
                  & L &  &    -0.7 &     0.9 &    -5.5 &     5.0 \\ 
  Cluster 10      & N &  &  2572.6 &  1845.8 &    95.5 & 10918.0 \\ 
                  & G &  &    16.5 &    14.9 &     0.1 &   123.6 \\ 
                  & L &  &    -0.6 &     0.9 &    -5.2 &     5.0 \\
 \hline
\end{tabular}
\end{table}

\begin{table}
\centering
\caption{\label{table:sup3} Estimation errors over process simulated plots for detection conditions \emph{full} visibility, \emph{centre} point visibility and \emph{any} visibility, and for estimators by us (\emph{HT}), Olofsson and Olsson (2018) (\emph{O \& O}), and Kuronen et al. (2019). Lowest (absolute) values for each case are bolded. The estimators O \& O and Kuronen et al. coincide in the centre point case.}
\begin{tabular}{llllrrrr}
  \hline
                 &        &                &  & \multicolumn{2}{c}{N}  & \multicolumn{2}{c}{G} \\ 
 \cline{5-6} \cline{7-8} \\
  Process         & Condition   & Estimator      &           & RMSE\% & ME\% & RMSE\% & ME\% \\ 
   \hline
Poisson         & full   & HT             &           &   \textbf{6.1} &   \textbf{0.0} &  \textbf{13.6} &   \textbf{0.3} \\ 
                  &        & O \& O        &           &  11.1 &  -7.5 &  16.1 &  -8.3 \\ 
                  &        & Kuronen et al. &           &   6.1 &   0.7 &  16.5 &   1.7 \\ 
                  &        & detected           &           &  28.3 & -21.5 &  34.0 & -23.3 \\ 
                  & centre & HT             &           &   4.8 &   \textbf{0.0} &   \textbf{7.8} &   \textbf{0.1} \\
                  &        & Kuronen et al.  &           &   \textbf{4.8} &   0.2 &   8.2 &   0.6 \\ 
                  &        & detected           &           &  20.3 & -15.1 &  23.8 & -16.0 \\ 
                  & any    & HT             &           &   3.4 &   \textbf{0.0} &   \textbf{5.0} &   \textbf{0.0} \\ 
                  &        & O \& O        &           &  12.0 &   8.5 &  15.2 &   9.6 \\ 
                  &        & Kuronen et al. &           &   \textbf{3.4} &   0.1 &   5.1 &   0.2 \\ 
                  &        & detected           &           &  11.9 &  -8.4 &  14.1 &  -8.8 \\ 
  Non-overlapping & full   & HT             &           &   \textbf{5.9} &   \textbf{0.4} &  \textbf{14.4} &   \textbf{0.5} \\ 
                  &        & O \& O        &           &  10.7 &  -7.2 &  16.6 &  -8.1 \\ 
                  &        & Kuronen et al. &           &   6.1 &   1.1 &  16.1 &   1.9 \\ 
                  &        & detected           &           &  27.9 & -21.2 &  34.1 & -23.1 \\ 
                  & centre & HT             &           &   \textbf{4.5} &   \textbf{0.3} &   \textbf{8.1} &   \textbf{0.5} \\
                  &        & Kuronen et al. &           &   4.5 &   0.6 &   8.1 &   1.0 \\ 
                  &        & detected           &           &  19.9 & -14.9 &  23.5 & -15.7 \\ 
                  & any    & HT             &           &   3.3 &   \textbf{0.0} &   5.4 &   \textbf{0.0} \\ 
                  &        & O \& O        &           &  12.1 &   8.6 &  15.1 &   9.7 \\ 
                  &        & Kuronen et al. &           &   \textbf{3.3} &   0.1 &   \textbf{5.1} &   0.2 \\ 
                  &        & detected           &           &  11.7 &  -8.3 &  14.1 &  -8.7 \\ 
  Gibbs 1        & full   & HT             &           &   8.0 &   4.3 &  12.7 &   5.0 \\ 
                  &        & O \& O        &           &   \textbf{7.0} &  \textbf{-4.0} &  \textbf{10.5} &  \textbf{-4.5} \\ 
                  &        & Kuronen et al. &           &   8.5 &   5.0 &  14.0 &   6.3 \\ 
                  &        & detected           &           &  24.3 & -18.4 &  28.9 & -20.1 \\ 
                  & centre & HT             &           &   \textbf{5.2} &   \textbf{2.1} &   \textbf{7.7} &   \textbf{2.4} \\ 
                  &        & Kuronen et al. &           &   5.3 &   2.4 &   8.0 &   2.9 \\ 
                  &        & detected           &           &  17.7 & -13.3 &  20.8 & -14.1 \\ 
                  & any    & HT             &           &   3.0 &   \textbf{0.3} &   \textbf{4.4} &   \textbf{0.3} \\ 
                  &        & O \& O        &           &  12.3 &   8.9 &  15.4 &  10.1 \\ 
                  &        & Kuronen et al. &           &   \textbf{3.0} &   0.4 &   4.4 &   0.4 \\ 
                  &        & detected           &           &  11.1 &  -7.9 &  12.9 &  -8.3 \\ 
  Gibbs 1.5      & full   & HT             &           &  14.3 &   9.5 &  39.5 &   6.0 \\ 
                  &        & O \& O        &           &   \textbf{3.6} &   \textbf{0.0} &  \textbf{36.7} &  \textbf{-3.6} \\ 
                  &        & Kuronen et al. &           &  14.5 &   9.8 &  40.5 &   7.2 \\ 
                  &        & detected           &           &  19.4 & -14.8 &  43.7 & -19.4 \\ 
                  & centre & HT             &           &   8.1 &   \textbf{4.9} &  \textbf{11.6} &   \textbf{5.1} \\ 
                  &        & Kuronen et al. &           &   \textbf{8.0} &   4.9 &  11.7 &   5.4 \\ 
                  &        & detected           &           &  14.0 & -10.8 &  17.7 & -11.7 \\ 
                  & any    & HT             &           &   3.9 &   1.8 &   4.3 &   \textbf{0.8} \\ 
                  &        & O \& O        &           &  14.7 &  10.7 &  15.7 &  10.8 \\ 
                  &        & Kuronen et al. &           &   \textbf{3.9} &   \textbf{1.7} &   \textbf{4.0} &   0.8 \\ 
                  &        & detected           &           &   8.6 &  -6.0 &  11.7 &  -7.3 \\ 
   \hline
\end{tabular}
\end{table}

\begin{table}
\centering
\caption{\label{table:sup4} Estimation errors over process simulated plots for detection conditions \emph{full} visibility, \emph{centre} point visibility and \emph{any} visibility, and for estimators by us (\emph{HT}), Olofsson and Olsson (2018) (\emph{O \& O}), and Kuronen et al. (2019). Lowest (absolute) values for each case are bolded. The estimators O \& O and Kuronen et al. coincide in the centre point case.}
\begin{tabular}{llllrrrr}
  \hline
                 &        &                &  & \multicolumn{2}{c}{N}  & \multicolumn{2}{c}{G} \\ 
 \cline{5-6} \cline{7-8} \\
  Process         & Condition   & Estimator      &           & RMSE\% & ME\% & RMSE\% & ME\% \\ 
   \hline
     Cluster 2       & full   & HT             &           &  17.2 &  -9.2 &  22.3 &  -9.8 \\ 
                  &        & O \& O        &           &  23.8 & -15.3 &  28.5 & -16.6 \\ 
                  &        & Kuronen et al. &           &  \textbf{16.3} &  \textbf{-8.2} &  \textbf{21.6} &  \textbf{-8.2} \\ 
                  &        & detected           &           &  40.7 & -28.1 &  47.3 & -30.3 \\ 
                  & centre & HT             &           &  12.0 &  -5.6 &  13.8 &  -5.5 \\ 
                  &        & Kuronen et al. &           &  \textbf{11.5} &  \textbf{-5.0} &  \textbf{13.3} &  \textbf{-4.7} \\ 
                  &        & detected           &           &  30.0 & -19.8 &  33.9 & -20.7 \\ 
                  & any    & HT             &           &   6.8 &  -2.1 &   7.8 &  -2.0 \\ 
                  &        & O \& O        &           &  11.7 &   6.4 &  15.0 &   7.5 \\ 
                  &        & Kuronen et al. &           &   \textbf{6.6} &  \textbf{-1.7} &   \textbf{7.7} &  \textbf{-1.5} \\ 
                  &        & detected           &           &  18.0 & -10.6 &  20.2 & -11.0 \\ 
  Cluster 4       & full   & HT             &           &  21.3 & -10.6 &  30.7 & -11.8 \\ 
                  &        & O \& O        &           &  27.4 & -16.4 &  36.2 & -18.3 \\ 
                  &        & Kuronen et al. &           &  \textbf{19.7} &  \textbf{-9.3} &  \textbf{30.1} &  \textbf{-9.6} \\ 
                  &        & detected           &           &  46.1 & -29.4 &  56.0 & -32.2 \\ 
                  & centre & HT             &           &  15.3 &  -6.9 &  19.7 &  -7.1 \\ 
                  &        & Kuronen et al. &           &  \textbf{14.4} &  \textbf{-6.0} &  \textbf{18.8} &  \textbf{-5.9} \\ 
                  &        & detected           &           &  35.0 & -21.2 &  41.0 & -22.4 \\ 
                  & any    & HT             &           &   9.3 &  -3.2 &  12.4 &  -3.5 \\ 
                  &        & O \& O        &           &  14.4 &   5.8 &  18.1 &   6.7 \\ 
                  &        & Kuronen et al. &           &   \textbf{8.8} &  \textbf{-2.5} &  \textbf{11.8} &  \textbf{-2.7} \\ 
                  &        & detected           &           &  22.3 & -11.9 &  26.5 & -12.7 \\ 
  Cluster 6       & full   & HT             &           &  20.6 &  -9.9 &  30.2 & -11.4 \\ 
                  &        & O \& O        &           &  27.9 & -16.1 &  36.6 & -18.0 \\ 
                  &        & Kuronen et al. &           &  \textbf{18.2} &  \textbf{-8.3} &  \textbf{29.0} &  \textbf{-9.1} \\ 
                  &        & detected           &           &  50.4 & -29.7 &  58.4 & -32.3 \\ 
                  & centre & HT             &           &  14.5 &  -6.4 &  19.0 &  -6.9 \\ 
                  &        & Kuronen et al. &           &  \textbf{13.1} &  \textbf{-5.4} &  \textbf{17.8} &  \textbf{-5.6} \\ 
                  &        & detected           &           &  38.7 & -21.5 &  43.1 & -22.7 \\ 
                  & any    & HT             &           &   8.7 &  -3.0 &  11.3 &  -3.1 \\ 
                  &        & O \& O        &           &  18.1 &   6.6 &  20.2 &   7.5 \\ 
                  &        & Kuronen et al. &           &   \textbf{8.1} &  \textbf{-2.3} &  \textbf{10.7} &  \textbf{-2.3} \\ 
                  &        & detected           &           &  25.3 & -12.4 &  27.8 & -12.7 \\ 
 \hline
\end{tabular}
\end{table}

\begin{table}
\centering
\caption{\label{table:sup5}  Estimation errors over process simulated plots for detection conditions \emph{full} visibility, \emph{centre} point visibility and \emph{any} visibility, and for estimators by us (\emph{HT}), Olofsson and Olsson (2018) (\emph{O \& O}), and Kuronen et al. (2019). Lowest (absolute) values for each case are bolded. The estimators O \& O and Kuronen et al. coincide in the centre point case.}
\begin{tabular}{llllrrrr}
  \hline
                 &        &                &  & \multicolumn{2}{c}{N}  & \multicolumn{2}{c}{G} \\ 
 \cline{5-6} \cline{7-8} \\
  Process         & Condition   & Estimator      &           & RMSE\% & ME\% & RMSE\% & ME\% \\ 
   \hline
  Cluster 8       & full   & HT             &           &  18.6 &  -7.7 &  24.9 &  -8.3 \\ 
                  &        & O \& O        &           &  26.0 & -14.2 &  31.2 & -15.5 \\ 
                  &        & Kuronen et al. &           &  \textbf{16.4} &  \textbf{-6.0} &  \textbf{21.7} &  \textbf{-5.8} \\ 
                  &        & detected           &           &  49.2 & -28.3 &  56.4 & -30.3 \\ 
                  & centre & HT             &           &  13.3 &  -5.0 &  16.6 &  -5.2 \\ 
                  &        & Kuronen et al. &           &  \textbf{12.0} &  \textbf{-4.0} &  \textbf{14.5} &  \textbf{-3.9} \\ 
                  &        & detected           &           &  37.9 & -20.5 &  42.8 & -21.5 \\ 
                  & any    & HT             &           &   8.0 &  -2.3 &  11.3 &  -2.6 \\ 
                  &        & O \& O        &           &  18.0 &   7.5 &  21.8 &   8.3 \\ 
                  &        & Kuronen et al. &           &   \textbf{7.5} &  \textbf{-1.7} & \textbf{10.0} &  \textbf{-1.8} \\ 
                  &        & detected           &           &  25.0 & -11.9 &  28.5 & -12.4 \\ 
  Cluster 10      & full   & HT             &           &  15.0 &  -6.4 &  24.1 &  -7.5 \\ 
                  &        & O \& O        &           &  22.5 & -13.0 &  30.4 & -14.9 \\ 
                  &        & Kuronen et al. &           &  \textbf{13.1} &  \textbf{-4.9} &  \textbf{21.7} &  \textbf{-5.2} \\ 
                  &        & detected           &           &  44.1 & -26.9 &  55.2 & -29.9 \\ 
                  & centre & HT             &           &  10.6 &  -4.0 &  16.6 &  -4.8 \\  
                  &        & Kuronen et al. &           &   \textbf{9.5} &  \textbf{-3.2} &  \textbf{15.1} &  \textbf{-3.6} \\ 
                  &        & detected           &           &  33.0 & -19.3 &  41.4 & -21.2 \\ 
                  & any    & HT             &           &   6.7 &  -1.9 &  11.0 &  -2.2 \\ 
                  &        & O \& O        &           &  15.6 &   7.6 &  21.2 &   8.6 \\ 
                  &        & Kuronen et al. &           &   \textbf{6.1} &  \textbf{-1.4} &   \textbf{9.9} &  \textbf{-1.5} \\ 
                  &        & detected           &           &  20.8 & -11.0 &  27.1 & -12.0 \\
 \hline
\end{tabular}
\end{table}

\begin{table}
\centering
\caption{\label{table:sup6} The percentages of simulated stem densities belonging to their approximate confidence intervals for detection conditions \emph{full} visibility, \emph{centre} point visibility and \emph{any} visibility.}

\begin{tabular}{lll rrr}
  \hline
 &  & & \multicolumn{3}{c}{interval}\\
\cline{4-6} \\
Data                 & Condition       &  & 90\% & 95\% & 99\% \\ 
   \hline
Field          & full   &          & 94.0 & 97.5 & 99.5 \\ 
                  & centre &          & 94.1 & 97.4 & 99.3  \\ 
                  & any    &          & 93.4 & 96.5 & 98.8 \\ 
Poisson         & full   &       & 90.0 & 94.9 & 98.7 \\ 
                  & centre &       & 89.9 & 94.5 & 98.7 \\ 
                  & any    &       & 90.5 & 94.9 & 98.4 \\ 
  Non-overlapping & full   &       & 90.7 & 95.7 & 99.0 \\ 
                  & centre &       & 91.5 & 95.7 & 98.6 \\ 
                  & any    &       & 91.0 & 95.0 & 98.4 \\ 
  Gibbs 1        & full   &       & 87.0 & 93.5 & 99.0 \\ 
                  & centre &       & 91.4 & 96.1 & 99.8 \\ 
                  & any    &       & 93.2 & 96.8 & 99.1 \\ 
  Gibbs 1.5      & full   &       & 63.6 & 74.3 & 88.4 \\ 
                  & centre &       & 81.2 & 89.3 & 97.4 \\ 
                  & any    &       & 81.5 & 90.3 & 98.8 \\ 
  Cluster 2       & full   &       & 60.2 & 67.0 & 76.8 \\ 
                  & centre &       & 68.5 & 75.1 & 84.2 \\ 
                  & any    &       & 80.0 & 85.2 & 92.1 \\ 
  Cluster 4       & full   &       & 57.0 & 64.1 & 74.2 \\ 
                  & centre &       & 64.3 & 70.8 & 79.6 \\ 
                  & any    &       & 73.9 & 79.5 & 87.2 \\ 
  Cluster 6       & full   &       & 60.2 & 66.2 & 76.8 \\ 
                  & centre &       & 68.1 & 74.1 & 82.8 \\ 
                  & any    &       & 75.5 & 81.8 & 88.9 \\ 
  Cluster 8       & full   &       & 67.8 & 74.1 & 83.6 \\ 
                  & centre &       & 73.2 & 78.9 & 87.1 \\ 
                  & any    &       & 79.7 & 85.5 & 92.1 \\ 
  Cluster 10      & full   &       & 70.6 & 77.0 & 85.5 \\ 
                  & centre &       & 75.7 & 81.2 & 89.0 \\ 
                  & any    &       & 81.2 & 85.4 & 92.4 \\ 
   \hline
\end{tabular}
\end{table}

\begin{table}
\caption{\label{table:sup7} The percentages of simulated basal areas belonging to their approximate confidence intervals for detection conditions \emph{full} visibility, \emph{centre} point visibility and \emph{any} visibility.}
\centering
\begin{tabular}{lll rrr}
   \hline
 &  & & \multicolumn{3}{c}{interval}\\
\cline{4-6} \\
  Data               &  Condition      &  & 90\% & 95\% & 99\% \\
   \hline
Field          & full   &          & 95.2 & 97.9 & 99.4 \\ 
                  & centre &          & 95.2 & 97.4 & 99.1 \\  
                  & any    &          & 94.2 & 96.5 & 98.6 \\  
Poisson         & full   &       & 89.4 & 94.4 & 98.0\\ 
                  & centre &       & 90.3 & 94.4 & 98.3 \\ 
                  & any    &       & 91.1 & 95.3 & 98.3 \\ 
  Non-overlapping & full   &       & 90.2 & 94.8 & 98.4 \\ 
                  & centre &       & 91.8 & 95.3 & 98.5 \\ 
                  & any    &       & 91.3 & 95.6 & 98.2 \\ 
  Gibbs 1        & full   &       & 86.9 & 92.9 & 98.9 \\ 
                  & centre &       & 91.3 & 95.7 & 99.3 \\ 
                  & any    &       & 92.7 & 96.5 & 98.8 \\ 
  Gibbs 1.5      & full   &       & 47.9 & 55.3 & 68.0 \\ 
                  & centre &       & 81.3 & 89.8 & 97.5 \\ 
                  & any    &       & 81.2 & 90.0 & 98.7 \\ 
  Cluster 2       & full   &       & 60.0 & 67.2 & 76.8 \\ 
                  & centre &       & 69.4 & 76.1 & 84.9 \\ 
                  & any    &       & 81.2 & 86.0 & 92.4 \\ 
  Cluster 4       & full   &       & 58.1 & 64.2 & 74.2 \\ 
                  & centre &       & 66.0 & 71.8 & 80.1 \\ 
                  & any    &       & 75.9 & 80.5 & 87.8 \\ 
  Cluster 6       & full   &       & 61.2 & 66.8 & 76.3 \\ 
                  & centre &       & 68.0 & 74.3 & 82.8 \\ 
                  & any    &       & 77.2 & 82.8 & 89.5 \\ 
  Cluster 8       & full   &       & 68.3 & 74.0 & 82.8 \\ 
                  & centre &       & 74.0 & 79.5 & 87.6 \\ 
                  & any    &       & 80.1 & 86.0 & 92.0 \\ 
  Cluster 10      & full   &       & 71.2 & 77.0 & 85.7 \\ 
                  & centre &       & 76.2 & 81.6 & 88.9 \\ 
                  & any    &       & 81.8 & 86.5 & 93.0 \\ 
   \hline
\end{tabular}
\end{table}

\begin{figure}
\centerline{\includegraphics[width=14cm]{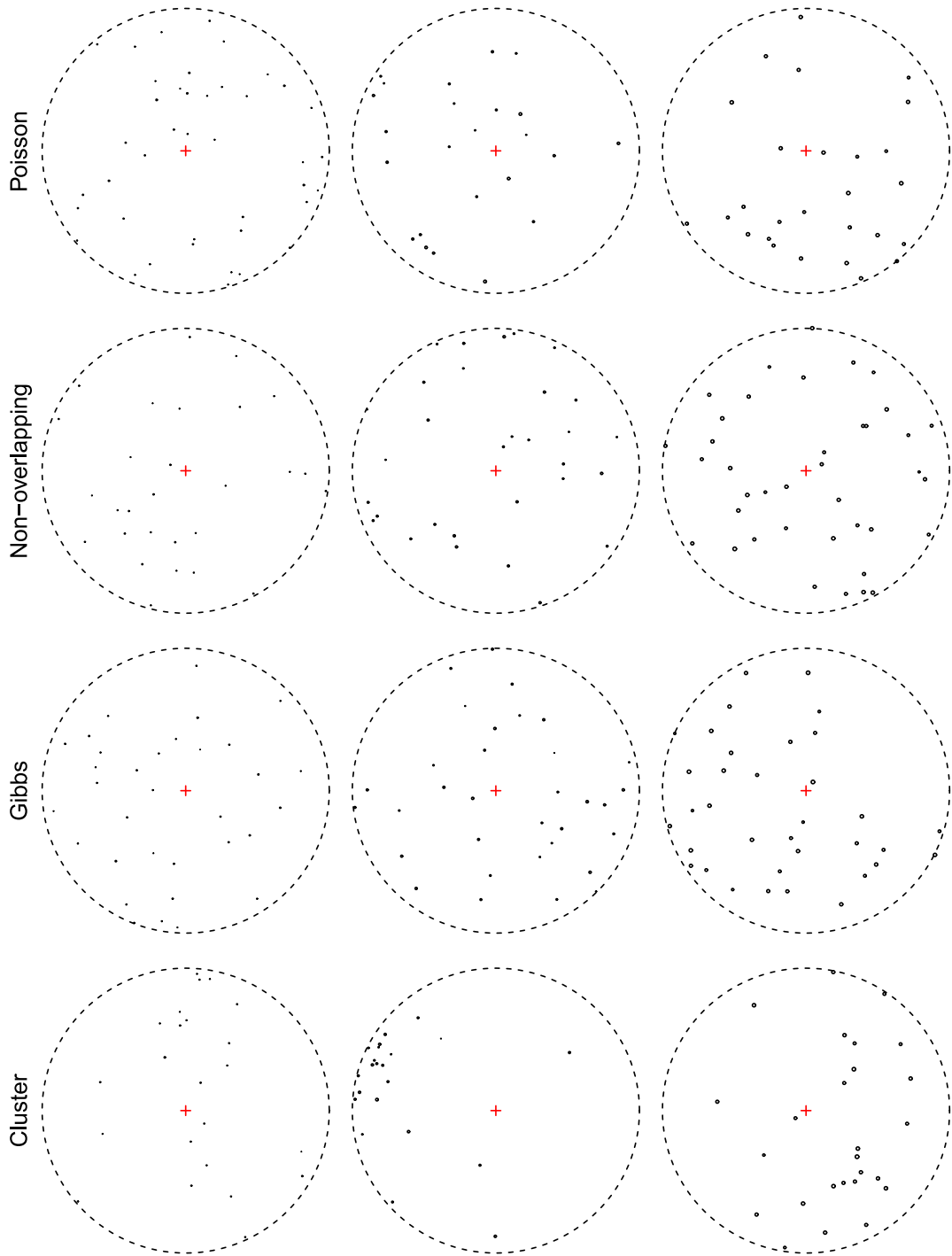}}
\caption{\label{fig:sup1} Three simulated plots from the Poisson process, Non-overlapping discs process, Gibbs hard core process, and Log-Gaussian Cox process that produces clustered patterns.}
\end{figure}

\end{document}